\documentclass {elsart}
\usepackage{amssymb}
\usepackage{graphicx}

\begin{document}
\begin{frontmatter}
\title{Evolution of Vocabulary on Scale-free and Random Networks }
\author [auth] {Alkiviadis Kalampokis},
\author[gi]{Kosmas Kosmidis} and
\author[auth]{Panos Argyrakis} 

\address [auth]{Department of Physics, University of Thessaloniki, 54124 Thessaloniki, Greece}
\address[gi]{Institut f\"ur Theoretische Physik III,
Justus-Liebig-Universit\"at, Giessen, Germany}

\date{08 -01-2007}

\begin{abstract}
We examine the evolution of the vocabulary of a group of individuals (linguistic agents) on a scale-free network, using Monte Carlo simulations and assumptions from evolutionary game theory. It is known that when the agents are arranged in a two-dimensional lattice structure and interact by diffusion and encounter, then their final vocabulary size is the maximum possible. Knowing all available words is essential in order to increase the probability to ``survive'' by effective reproduction. On scale-free networks we find a different result. It is not necessary to learn the entire vocabulary available. Survival chances are increased by using the vocabulary of the ``hubs'' (nodes with high degree). The existence of the ``hubs'' in a scale-free network is the source of an additional important fitness generating mechanism.

\end{abstract}



\end{frontmatter}

\section {Introduction}
Human language and its evolution has recently become an attractive subject within the interdisciplinary scientific community \cite{Nowak1,Strog,Sch1,Sch2,Sch3,Kos1,Kos2,Baron,Kos3,Schwam1,Schwam2,Meyort,Patr,Mira,Pina,viv1,viv2,Hav1,Can,baxt,Wich}.
The reason for this interest is a natural consequence of the rapid advances in the understanding and modeling of complex systems\cite{Stauf1}. Statistical physics, and mainly computational statistical physics, has proven to be quite effective in the study of systems of many interacting atoms and in the description of several complex phenomena associated with these interactions, even though the interacting units are no longer atoms or elementary particles but biological species\cite{Droz}, human beings \cite{Penna,Gallos} or even financial tools, such as stocks \cite{Mant}. It was also realized that human language, a traditionally qualitative subject of study, fits adequately in the above quantitative framework. Several aspects of language have been studied by different groups. The main focus is on language learning and its evolution \cite{Nowak1,Kos2,Baron},
on the quantification of language characteristics (for example, the famous Zipf law) and their explanation from first principles \cite{Kos2,Kos3,Can}, and on language competition between two \cite{Strog,Kos1,Schwam1,Patr} or more languages \cite{Sch1,Sch2,Sch3,Schwam2,Meyort,viv1,viv2}. In several of the above studies, for example \cite{Sch1,Kos1}, the authors assume that language is learned by linguistic agents that move on a regular two-dimensional lattice (a surface) and interact with each other. The effect of the surface topology or possible disorder is not taken into account. It has been, however, recently understood that a lattice topology may in several cases be an inadequate substrate for the description of social interactions. In many cases, a better description is achieved if one takes into account that social systems may be represented as graphs (networks), i.e. as collections of nodes (representing individuals) which are connected together if the individuals represented by these nodes know each other. Social networks have a structure similar to a scale-free network \cite{Amaral,NWS,Lij}, which is a graph whose degree (the number of edges that emanates from a node) distribution follows a power law, $P(k)\sim k^{-\gamma}$, and have attracted considerable interest\cite{Amaral,NWS,Lij,BAL,AJB,Newm,Cohen,Riitta}.
\\During most of human history, words were learned by individuals through discussions with those close to the learner. In the last decades, however, ``modern'' technologies have changed this situation. We now have mobile phones, e-mail accounts, web cameras and communication with even the most remote acquaintance is not only possible but has become a rather easy task. Motivated by this fact, we study language evolution on scale-free network structures. We use Monte Carlo simulations and assumptions from evolutionary game theory in order to evaluate the way the network topology affects the vocabulary size of a group of individuals. In this way we hope to get an insight on how ``modern'' technologies may affect language and its evolution. Monte Carlo simulations on regular geometries have shown \cite{Kos1} that when agents are arranged in a two-dimensional lattice structure, their final vocabulary size is the maximum possible. Scale-free networks, in contrast to regular lattices, are characterized by the existence of nodes with very high degree (``hubs'') and our results indicate that these hubs have an important impact on the vocabulary size. Network theory has been used in the past to study the structure of language (\cite{Can} and references therein). Here, however, we use it in order to study vocabulary evolution, independently of any linguistic structure.

\section{Model and Methods}
We build a computational model to study the time evolution of the vocabulary known by species which interact with each other. The model is in several aspects similar to a previous model used in \cite{Kos1} for describing language evolution on a square lattice topology. In order to determine the effect of the network topology to the vocabulary learning characteristics, we used scale free networks with $N=10000$ nodes and with $\gamma=2.0, 2.5$ and $3.0$, and Erd\"os - R\'enyi networks also consisting of $N=10000$ nodes and connectivity probability $\rho = 0.002$. Each node is always occupied by one agent. The language that these agents have consists of 10 words. Thus, each agent possesses a maximum vocabulary of 10 words. The number of words that a given agent knows at any time is not constant, since there are mechanisms to learn and to forget words, which will be explained in the next paragraphs.
Each agent has a number of attributes that characterize its behavior. The first is the vocabulary $V$ which consists of an array of 10 elements. An element has a value of 1 if the corresponding word is known to the agent or 0 if the word is unknown. Initially, each agent has a vocabulary that consists of 5 words, chosen randomly out of the 10 possible words. The second is the fitness, $f$, which determines the probability of each agent to reproduce. The initial fitness has a value of zero, and agents can gain fitness through successful communication.
All agents take part in the following activities: Communication, reproduction and mortality.
\\1) \emph{Communication}: An agent $i$ is chosen randomly and given the possibility to communicate with one of its neighbors, $j$. As neighbors we consider the nodes with which the specific node is connected. This communication confers fitness to both agents ($i$ and $j$) according to the number of words agent $i$ has in common with agent $j$ with which it communicates. Specifically:
\\a) The payoff for the interaction is equal to the number of words $i$ and $j$ have in common (e.g., three common words means a payoff of 3). This payoff value is added to the fitness of each agent, as a reward for successful communication.
\\b) Learn-forget process: Every word in the vocabulary is examined and if agent $i$ does not know a word which is known to $j$ then there is a probability $p_{L}$ that $i$ will learn it from $j$. If this specific word is learned, then the corresponding vocabulary array element will turn from 0 to 1. However, there is also a probability $p_{F}$ that $j$ will forget this word not known to $i$. The same rules apply for words which are known to $i$ and unknown to $j$. Thus, words that are unknown to the majority of the population have increased probability of being lost from the language. 
\\2) \emph{Reproduction}: There is a probability $p_{r}$ that a reproduction event will take place. The selection of the agent to be reproduced is not random but proportional to the agent's fitness. This means that agents with large fitness have a higher probability for reproduction. Each agent's probability for reproduction is given by the formula:

\begin{equation}
p_{i}=\frac{f_{i}}{ \sum_{i} f_{i}} 
\end{equation} 

where $f_{i}$ is the fitness of agent $i$ and this sum is over all agents.  The fact that we normalize over the total fitness implies that there is information available to all agents about the fitness status of their society. Since all the sites (or nodes) of our space are occupied the offspring will have to be born in an already occupied site, replacing the previous inhabitant. In this way the reproduction and mortality procedures of the model are combined in one action as opposed to the model used in \cite{Kos1}. We have two choices for the selection of the site in which the offspring will be born, and live thereafter. The first is to put it in one of the neighboring sites of the parent, which seems quite rational, for the child to ``live'' near its parents. The second is to choose a random site and place it there. Although both models give qualitatively the same results, there are numerical differences. In the ``local'' model we observed more fluctuations, while in the ``global'' model these fluctuations were not persistent and soon smoothed out. For this reason in the current text we will present only the ``global'' model, where the offspring takes a randomly chosen site in the network. The next choice to be made concerns the amount of fitness that the offspring will inherit. For simplicity, we assume that the child inherits the fitness of the parent. Thus, the offspring begins its life having the same amount of fitness its parent has, without affecting the parents fitness. All offspring carry the full vocabulary of the parent.
\\3) After each cycle of communication and reproduction, time is incremented by $1/N$, where $N$ is the total number of agents in the lattice. Thus, one time unit or Monte Carlo Step (MCS) statistically represents the time necessary for each agent to execute the communication-reproduction cycle once. The simulation continues until a predefined total time is reached.
For statistical purposes we average our results over a large number of realizations, typically  1000, in this work. In most cases the time evolution of the system is followed up to 100000 MCS, but since the system reaches a state of equilibrium much sooner the data we show here are limited to 20000 MCS. In all simulation results presented in the present manuscript we have used the values $p_{L}=0.1$, $p_{F}=0.1$.

\section{Results}
\subsection {No Reproduction, $p_{r}=0$.}
First we studied the behavior of our system when no reproduction takes place. In this case only the communication process is active and the agents can pass linguistic information to their neighbors. One expects that after sufficient time this process would stop since the system would reach a steady state where all the agents have acquired exactly the same vocabulary. We studied  scale free networks, with  $\gamma=2.0, 2.5$ and $3.0$ and random networks with connectivity $\rho = 0.002$. Both netowrks consisted of $N=10000$ nodes,
The results for the mean number of words known by each agent have shown that there is no real change in the number of words known by the agents. Instead, there is only some very small fluctuation around the number of words that the inhabitants of the system know at the start of the simulation. This means that there is no tendency for ``knowledge'' to spread around the network. This can be expected since if there is no reproduction, the knowledge of many words is \emph{not} an evolutionary advantage. Moreover, there are no newborn agents who spread around the network, spreading also their vocabulary. This model is, therefore, quite static, both in its rules and in the results that we get. This result is similar to the one obtained in \cite{Kos1} for the case of no reproduction on a lattice and agrees with intuition.

\subsection {Reproduction, $p_{r}\neq 0$.}
In a previous work \cite{Kos1}, it was shown that when identical (i.e having the same initial fitness) linguistic agents are allowed to move on a lattice, to learn and forget words as described in the methods section and to reproduce with a probability $p_{r}\neq 0$ then the final state of the system is one where all agents have learned all possible words. To be precise, if the agents move on a lattice and initially they know on average 5 words (i.e if the have 5 digits equal to one in their vocabulary array $V$ which has size equal to 10) we will end up in a situation where all agents know an average of 10 words. This is reasonable as language is a fitness generating mechanism and thus, knowing many words is essential for survival. 
The ``survival of the fittest'' implies that in order to survive, one has to know everything that is available and this is verified by simulations. On a scale-free network, however, the situation is different. 
\begin{figure}
\begin{center}
\includegraphics[height=8.5 cm,width=8.5cm]{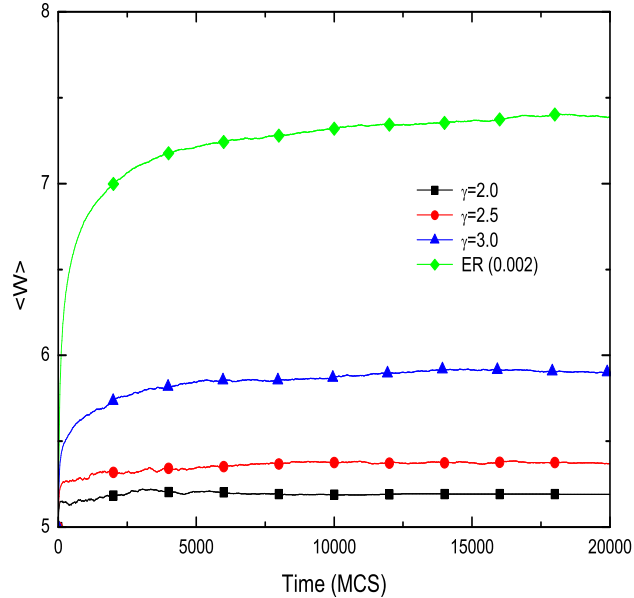}
\end{center}
\caption{Mean number of words, $<W>$, known by each agent, vs. time, for scale free networks consisting of $N=10000$ nodes, for $\gamma=2.0, 2.5, 3.0$ and random networks consisting of $N=10000$ nodes, with connectivity $\rho =0.002$, for $p_{r} = 0.1$. The results are the averages of 1000 realizations.}
\label{fig1}
\end{figure}
In Figure \ref{fig1}, we plot the mean number of words, $<W>$, known by each agent as a function of time, for scale free networks consisting of $N=10000$ nodes, for $\gamma=2.0, 2.5, 3.0$ and random networks consisting of $N=10000$ nodes, with connectivity $\rho = 0.002$ for $p_{r} = 0.1$. It is obvious, especially in the case of networks with $\gamma=2$ that the number of words is quite below the maximum vocabulary size. The reason for this is that in such a network there are several nodes with very high degree (hubs). A hub has many neighbors and, consequently, it can communicate with many other nodes and  drasticly increase its fitness. This can be easily understood with the example of an ``extreme'' case. Consider a star-like network of say $N=100$ nodes. There is one central node with degree $k=99$, i.e. 
a hub, which is connected to all other nodes. The remaining 99 nodes have $k=1$, thus, they are connected to the hub only. In one time step each node is on average selected once and then it selects randomly one of its neighbors to communicate with. In this extreme case, at the first time step, each node has one chance to communicate, except the hub that has 99 chances because it is the only neighbor that a randomly selected node has. Thus, the hubs are in an advantageous position and gain fitness quickly. Then, they are favored in reproduction and finally their vocabulary dominates the system.
The existence of hubs has as a result that a node can gain fitness not only by knowing many words, but also by knowing just the words that are known to the hubs. There is, thus, a new fitness generating mechanism associated with the existence of the hubs. The effect is more profound for networks with $\gamma=2$, where there are a lot of large hubs and less evident for higher $\gamma$ values where the hubs are fewer.
It is also important that the offspring replaces a randomly chosen node, since thus it favors the spreading of the vocabulary of the fittest nodes, which in this case is the vocabulary of the hubs.
\\The case of the random networks is quite different, since here we observe a significant increase in the number of words the agents know, although they are still lower than those of the square lattice. Thus, random networks are between the two cases, showing the effect of scale free networks but in a much lesser extent, a fact we can safely assume is due to the lack of big hubs.

\begin{figure}
\begin{center}
\includegraphics[height=8.5 cm,width=8.5cm]{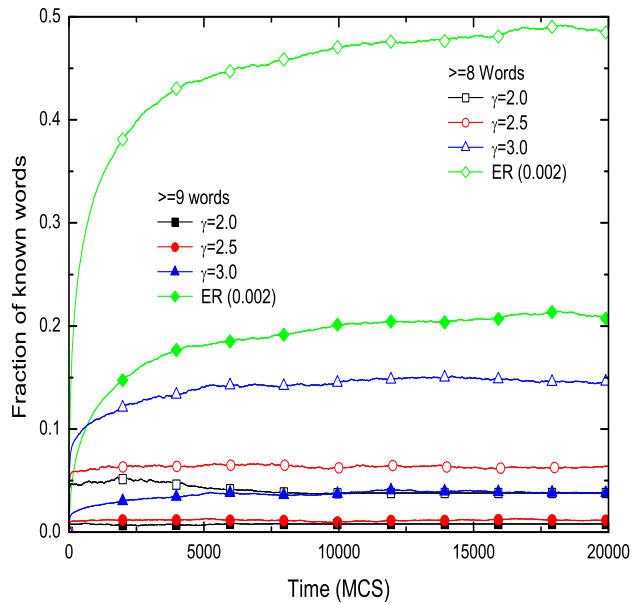}
\end{center}
\caption{The fraction of agents that know 8 or more words and 9 or more words, vs. time, for scale free networks consisting of $N=10000$ nodes, for $\gamma=2.0, 2.5, 3.0$ and random networks consisting of $N=10000$ nodes, with connectivity $\rho = 0.002$ and $p_{r} = 0.1$. The results are the averages of 1000 realizations.}
\label{fig2}
\end{figure}

In Figure \ref{fig2} we plot the fraction of the nodes that know 8 or more words and 9 or more words, as a function of time. Since now knowledge of more words is an evolutionary advantage we see that this fraction is much larger than in the non-reproduction case we saw in the previous paragraph. It is, however, lower than the values expected for a lattice topology and we can also observe that low $\gamma$ values are associated with lower fractions of agents with ``rich vocabulary'', in agreement with what we have previously mentioned. 

\subsection {Reproduction and language competition, $p_{r}\neq 0$.}
We have also simulated language competition between two interacting species on a scale free network using the algorithm described in detail in Ref \cite{Kos1}.  The main difference now is that we have two species speaking two different languages and instead of starting with a random vocabulary knowledge, half of the population knows perfectly one language and the other half knows perfectly a completely different language. In this case the maximum possible vocabulary size is 20 (there are two ``languages'' that have 10 words each) and we assume that the child inherits 80\% of the father's fitness \cite{koskalarg}.
It is obvious that inspite of the differences between the two algorithms, the different initial conditions on the word distribution and the difference in the fitness amount passed from one generation to the next, we can still observe that for scale-free networks with $\gamma=2$ the total number of words that are finally known by the nodes is significantly less than for the lattice case. This is an indication that the role of the hubs is significant for the propagation and learning of new words and that this fact does not strongly depend on the specific details of the models.

\begin{figure}
\begin{center}
\includegraphics[height=8.5 cm,width=8.5cm]{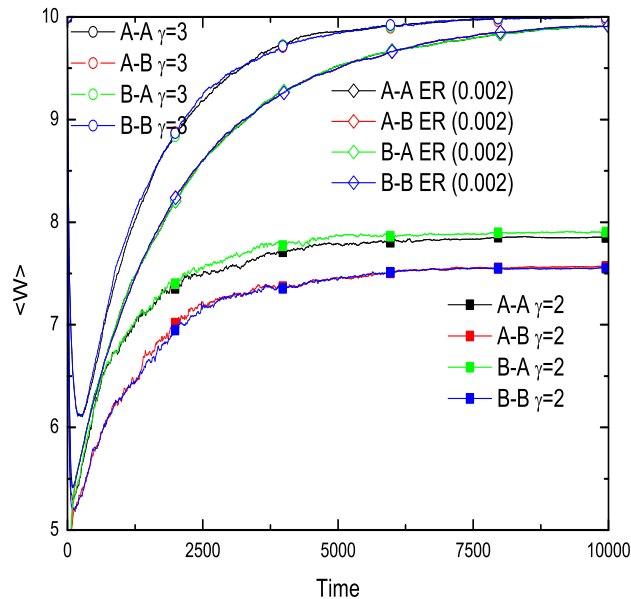}
\end{center}
\caption{Plot for the two language model, of the average number of words known to an agent vs time for scale free networks consisting of $N=10000$ nodes with $\gamma=2.0, 3.0$ and random networks consisting of $N=10000$ nodes with connectivity $\rho = 0.002$, $p_{r} = 0.1$ and the assumption that the child inherits 80\% of the fathers fitness. The initial concentration is $c=0.15$ for both A and B species}
\label{fig3}
\end{figure}

\section{Conclusions}
We have studied the evolution of the vocabulary of a group of individuals on a scale-free network. We have demonstrated that there is an important difference in this case, compared to the case where the individuals are regularly distributed on a lattice or even with the case where they are allowed to perform random walks on a lattice. On a lattice structure, the final vocabulary size of the individuals is the maximum possible. Knowing everything is essential in order to increase the probability to ``reproduce''. On scale-free networks, however, the reproduction probability is considerably increased by using the vocabulary of the ``hubs''. This result indicates that the existence of the ``hubs'' in a scale-free network is the source of an additional important fitness generating mechanism and may have profound and unexpected impact on the evolutionary dynamics of a system. 

\section*{Acknowledgements}
We would like to thank Dr. L.K. Gallos for fruitful discusions. This work was partially supported by the Hellenic Ministry of Education, via PYTHAGORAS project.

\end{document}